\documentstyle{article}

\font\tenrm=cmr10
\font\tenit=cmti10
\font\elevenbf=cmbx10 scaled\magstep 1
\font\elevenrm=cmr10 scaled\magstep 1
 1

\renewenvironment{thebibliography}[1]
 { \tenrm
 \baselineskip=10pt
   \begin{list}{\arabic{enumi}.}
    {\usecounter{enumi} \setlength{\parsep}{0pt}
     \setlength{\itemsep}{3pt} \settowidth{\labelwidth}{#1.}
     \sloppy
    }}{\end{list}}

\begin{document}
\begin{center}{\elevenbf
MULTIBARYONS WITH STRANGENESS, CHARM AND BOTTOM
\\}
\vglue 0.5cm
{\tenrm V.B.Kopeliovich$^*$ and  W.J.Zakrzewski$^{**}$ \\}
{\tenit $^*$Institute for Nuclear Research of the Russian Academy of
Sciences, Moscow 117312, Russia\\
and \\$^{**}$Department of Mathematical Sciences, University of Durham,
Durham DH1 3LE, UK\\}
\vglue 0.5cm
\end{center}
{\rightskip=3pc
\leftskip=3pc
\tenrm\baselineskip=12pt
\noindent
Static properties of multiskyrmions with baryon numbers up to $8$ are
calculated, including momenta of inertia and sigma-term. The calculations 
are based on the recently suggested $SU(2)$ rational map ansaetze.
The spectra of baryonic systems with strangeness, charm and bottom are 
considered within a ``rigid oscillator" version of the bound state 
soliton model. The binding energies estimates are made for the states
with largest isospin which can appear as negatively charged nuclear
fragments, as well as for states with zero isospin - light fragments of 
``flavoured" nuclear matter. Our results  confirm the previously made
observation that baryonic systems with charm or bottom 
quantum numbers have more chance to be stable with respect to strong 
interactions than strange baryonic systems. 
 \vglue 0.2cm}
 \vglue 0.2cm
\baselineskip=13pt
\elevenrm
\section{Introduction}
The topological soliton models, and the Skyrme model among them \cite{1}, 
are attractive
because of their simplicity and the possibility that they may describe well
 various properties of low energy baryons. The models of this kind provide
also a very good framework within which to investigate the possibility
of the existence of nuclear matter fragments with unusual properties, 
such as flavour being different from
$u$ and $d$ quarks. 
In addition to being important by itself, this issue can have
important consequences in astrophysics and cosmology. In particular, the
formation and subsequent decay of such fragments could 
be important in the early stages of the evolution of the Universe. 
 It is well
known that the relativistic many-body problems cannot be solved directly 
using the existing 
methods, and the chiral soliton approach may allow us to overcome
some of these difficulties.

The description of skyrmions with large baryon numbers has been perceived
as being  complicated because the explicit form of the fields has been not
known.
A recent observation \cite{2} that the fields
of the $SU(2)$ skyrmions can be 
approximated accurately by rational map ansaetze giving the values of 
masses close to their precise values, has simplified considerably the task
of such studies.
Similar ansaetze have also been recently presented for $SU(N)$ skyrmions
(which are not embeddings of $SU(2)$ fields)\cite{3}.

In this paper we use the $SU(2)$ rational map 
ansaetze as the starting points for the calculation of static
properties of bound states of skyrmions necessary for their quantization
in the $SU(3)$ collective coordinate  space.
The energy density of the  $B=3$ configuration has tetrahedral symmetry, 
of $B=4$ - the octahedral
(cubic) one \cite{4}, of $B=5$ - $D_{2d}$-symmetry, of $B=6$ - $D_{4d}$, 
of $B=7$ - dodecahedral symmetry, and of $B=8$ - $D_{6d}$ - symmetry 
\cite{5,2}, etc.

The minimization, with the help of a 3-dimensional variational program \cite{6},
lowers the energies of these configurations by few hundreds of $Mev$
and shows that they become local minima in the $SU(3)$ configuration space.
The knowledge of the so-called ``flavour" moment of inertia and the $\Sigma$-term allows us then to estimate the flavour excitation energies.
The mass splittings of the lowest states with different values
of strangeness, charm or bottom are calculated within the rigid oscillator
version of the bound state approach. The binding energies of baryonic systems
($BS$) with different values of flavours are also estimated.

To reduce theoretical uncertainties  we consider 
the differencies between the binding energies of $BS$ with flavour $F$ and
the ground
state for each value of $B$. These ground states are the deuteron for $B=2$,
the isodoublet $^3H$-$^3He$ for $B=3$, $^4He$ for $B=4$, etc.
These differencies, being free of many uncertainties, in particular of the
poorly known loop corrections to classical masses, show  the tendency
of the flavoured $BS$ to be more bound than the $(u,d)$ ground states (for heavy
flavours), or to be less bound, as for strangeness. Of course, it is an
assumption that the ground states of multiskyrmions correspond to ordinary
nuclei. However, this is a natural assumption if we believe that effective
field theories describe nature.

In the next Section the static characteristics of multiskyrmions are described.
Flavour excitation energies and zero mode corrections to the energies of 
multibaryons are considered in Section 3. Section 4 contains estimates for the
binding energies of baryonic systems with different values of flavours, and
our conclusions are given in Section 5.
\section{Static characteristics of multiskyrmions}
We consider here simple $SU(3)$ extensions of the Skyrme model \cite{1}: we 
start with $SU(2)$ skyrmions (with flavour 
corresponding to $(u,d)$ quarks) and extend them to various $SU(3)$ groups, 
such as, $(u,d,s)$, $(u,d,c)$, or $(u,d,b)$.

We take the Lagrangian of the Skyrme model which, in its well known form,
depends on parameters $F_{\pi}, \; F_D$ and $e$ and
can be written in the following way \cite{7}:
$${\cal L} =  \frac{F_\pi^2}{16}Tr l_{\mu} l^{\mu} + {1 \over 32e^2}
Tr [l_\mu,l_\nu]^2 +\frac{F_\pi^2m_\pi^2}{16} Tr(U+U^{\dagger}-2) + $$
$$ +\frac{F_D^2m_D^2-F_\pi^2m_\pi^2}{24}Tr(1-\sqrt{3}\lambda_8)(U+U^{\dagger}-2)+
 \frac{F_D^2-F_\pi^2}{48} Tr(1-\sqrt{3}\lambda_8)(Ul_\mu l^\mu +
l_\mu l^\mu U^{\dagger}).\eqno (1) $$
Here $U \in SU(3)$ is a unitary matrix incorporating chiral (meson) fields, and
$l_\mu=U^{\dagger} \partial _\mu U$. In this model $F_\pi$ is fixed 
at the physical value: $F_\pi$ = $186$ Mev and $M_D$ is the mass of $K, \, D$ or
$B$ meson. The ratios $F_D/ F_\pi$ are known to be $1.22$ and $1.7 \pm
0.2$ for, respectively, kaons and $D$-mesons.

The flavour symmetry breaking $(FSB)$ in the Lagrangian is of
 the usual form, and is sufficient to describe the mass 
splittings of the octet and decuplets of baryons
within the collective coordinate quantization approach \cite{7}. 

The Wess-Zumino term, to be added to the action, which can be written as a 5-dimensional
differential form \cite{8} plays an 
important role in the quantization procedure:
$$ S^{WZ}= \frac{-iN_c}{240 \pi^2}\int_\Omega d^5x\epsilon^{\mu\nu\lambda\rho
\sigma} Tr (l_\mu l_\nu l_\lambda l_\rho l_\sigma), \eqno (2) $$
where $\Omega $ is a 5-dimensional region with the 4-dimensional space-time 
as its boundary and $l_\mu = U^\dagger \partial_\mu U$.
Action $(2)$ defines important topological properties of skyrmions, but it 
does not contribute to the static masses of classical configurations  \cite{8,9}.
Variation of this action can be presented as a well defined contribution
to the Lagrangian (integral over the 4-dimensional space-time).

We begin our calculations, however, with $U \in SU(2)$.
The classical mass of $SU(2)$ solitons, in the most general case, depends on $3$
profile functions: $f, \, \alpha$ and $\beta$ and is given by
$$ M_{cl} = \int \biggl\{ \frac{F_\pi^2}{8} \bigl[\vec{l}_1^2+\vec{l}_2^2+
\vec{l}_3^2 \bigr] + {1 \over 2e^2} \bigl[ [\vec{l}_1\vec{l}_2]^2 +
[\vec{l}_2\vec{l}_3]^2 +[\vec{l}_3\vec{l}_1]^2 \bigr] + 
 {1 \over 4} F_\pi^2m_\pi^2 (1-c_f)\biggr\} d^3r \eqno (3) $$

Here $\vec{l}_k$ are the $SU(2)$ chiral derivatives defined by 
$U^{\dagger}\vec{\partial}U = i \vec{l}_k \tau_k$, where $k=1,2,3$.
The general parametrization of $U_0$ for an $SU(2)$ soliton we use here 
is given by
$U_0 = c_f+s_f \vec{\tau}\vec{n}$ with $n_z=c_{\alpha}$, $n_x=s_{\alpha}
c_{\beta}$, $n_y=s_{\alpha}s_{\beta}$, $s_f=sinf$, $c_f=cosf$, etc.
For the rational map ansatz we are using here as starting configurations,
$$n_x=\frac{2 Re\, R(\xi)}{1+|R(\xi)|^2}, \qquad
n_y=\frac{2 Im\, R(\xi)}{1+|R(\xi)|^2}, \qquad
n_z=\frac{1-|R(\xi)|^2}{1+|R(\xi)|^2}, $$
where $R(\xi)$ is a ratio of polynomials of the maximal power $B$ of the variable
$\xi=tg(\theta /2)exp(i\phi)$, $\theta$ and $\phi$ being polar and azimuthal
angles defining the direction of the radius-vector $\vec{r}$. The explicit
form of $R(\xi)$ is given in \cite{2} for different values of $B$.

The ``flavour" moment of inertia plays a very important role in the procedure
of $SU(3)$ quantization \cite{10}-\cite{18}, see formulas $(9),(10)$ below, 
and for arbitrary $SU(2)$ skyrmions is given by \cite{17,19}:
$$ \Theta_F = {1 \over 8} \int (1-c_f)\bigl[ F_D^2 + {1 \over e^2}
 \bigl( (\vec{\partial}f)^2 +s_f^2(\vec{\partial}\alpha)^2 +
s_f^2s_{\alpha}^2(\vec{\partial}\beta)^2\bigr) \bigr] d^3\vec{r}. \eqno (4a) $$
It is simply connected with $\Theta_F^{(0)}$ of the flavour symmetric case
($F_D=F_\pi$):
$$\Theta_F=\Theta_F^{(0)}+(F_D^2/F_\pi^2-1) \Gamma/4, \eqno (4b) $$
with $\Gamma$  defined in $(5)$ below.
The isotopic momenta of inertia are the components of the corresponding
tensor of inertia. They have been discussed in many papers, see e.g. 
\cite{9}-\cite{12}, so, we will not present them here. For majority of
multiskyrmions we discuss, this tensor of inertia is close to the unit
matrix multiplied by the isotopic moment of inertia $\Theta_T$.
This is exactly the case for $B=1$ and, to within 
a good accuracy, for $B=3,\,7 $. Considerable deviations take place for the
torus with $B=2$, and smaller ones for $B=4,\,5, \,6$ and $8$,
see {\bf Table 1}.
The quantity $\Gamma$ (or $\Sigma$-term), which defines the contribution of the 
mass term to the classical mass of solitons, and $\tilde{\Gamma}$ are used 
directly in the quantization procedure. They are given by:
$$ \Gamma = \frac{F_{\pi}^2}{2} \int (1-c_f) d^3\vec{r}, \qquad
  \tilde{\Gamma}={1 \over 4} \int c_f
 \bigl[ (\vec{\partial}f)^2 +s_f^2(\vec{\partial}\alpha)^2 +
s_f^2s_{\alpha}^2(\vec{\partial}\beta)^2 \bigr] d^3\vec{r}. \eqno (5)$$
The following relation can also be established:
$\tilde{\Gamma}=2(M_{cl}^{(2)}/F_\pi^2-e^2\Theta_F^{Sk})$, where 
$M_{cl}^{(2)}$
is the second-order term contribution to the classical mass of the soliton,
and $\Theta_F^{Sk}$ is the Skyrme term contribution to the flavour moment of
inertia.
The calculated masses of solitons, momenta of inertia $\Theta_F$, $\Theta_T$, 
$\Gamma$ or $\Sigma$-term and $\tilde{\Gamma}$ are presented
 in {\bf Table 1}.
\vspace{2mm}
\begin{center}
\begin{tabular}{|l|l|l|l|l|l|l||l|l|l|l|l|l|l|l|}
\hline
 $B$  &$M_{cl}$& $\Theta_F^{(0)}$ & $\Theta_T$&$\Theta_{T,3}$&$\Gamma$&
$\tilde{\Gamma}$&$\omega_s$&$\omega_c$ &$\omega_b$&$c_s$&$c_c$&$c_b$&
$\bar{c}_s$&$ \bar{c}_c$ \\
\hline
$1$&$1.702 $&$2.04$&$5.55$&$5.55$&$4.83$&$15.6$&$0.309$ &$1.542 $  &$4.82$
&$0.28$&$0.27$&$0.52$&$0.54$&$0.91$\\
\hline
$2$&$3.26 $&$4.18$&$11.5$&$7.38$&$9.35$&$22$&$0.293$ &$1.511 $&$4.76$ 
&$0.27$&$0.24$&$0.49$&$0.53$&$0.90$\\
\hline
$3$&$4.80 $&$6.34$&$14.4$&$14.4$&$14.0$&$27$&$0.289$ &$1.504 $&$4.75$ 
&$0.40$&$0.37$&$0.58$&$0.60$&$0.92$\\
\hline
$4$&$6.20 $&$8.27$&$16.8$&$20.3 $&$18.0$&$31 $&$0.283$ &$1.493$&$4.74 $
&$0.47$&$0.44$&$0.62$&$0.64$&$0.92$ \\
\hline
$5$&$7.78$ &$10.8$&$23.5 $&$19.5$&$23.8$&$35 $&$0.287 $&$1.505$&$4.75 $ 
&$0.42$ &$0.40$&$0.60$&$0.62$&$0.92$ \\
\hline
$6$&$9.24 $&$13.1$&$25.4$&$27.7$&$29.0$&$38 $&$0.287 $&$1.504 $&$4.75 $ 
&$0.48$&$0.46$&$0.63$&$0.67$&$0.93$ \\
\hline
$7$&$10.6 $&$14.7$&$28.9$&$28.9$&$32.3$&$44 $&$0.282 $&$1.497 $&$4.75 $ 
&$0.48$&$0.46$&$0.64$&$0.66$&$0.93$\\
\hline
$8$&$12.2 $&$17.4$&$33.4$&$31.4$&$38.9$&$47$&$0.288 $&$1.510 $&$4.79 $
&$0.49$&$0.47$&$0.64$&$0.67$&$0.93$\\
\hline
\end{tabular}
\end{center}
\vspace{2mm}
{\baselineskip=10pt 
\tenrm
{\bf Table 1.} Characteristics of the bound states of skyrmions
with baryon numbers up to $B=8$. The classical mass of solitons $M_{cl}$ is 
in $Gev$, momenta of inertia $\Theta_F, \, \Theta_T$ and $\Theta_{T,3}$,
$\Gamma$ and $\tilde{\Gamma}$ - in $Gev^{-1}$, the excitation frequencies 
for flavour $F$, $\omega_F$ in $GeV$. 
 $c_{s,c,b}$ and $\bar{c}_{s,c}$ are the hyperfine splitting constants for
multibaryons defined in Eq.$(21)$. The constant $\bar{c}_b$ is close to 
$0.99$ for all $B$ and is not included into the Table.
The parameters of the model
$F_{\pi}=186 \, MeV, \; e=4.12$. The accuracy of calculations is better
than $1\%$ for the masses and few $\%$ for other static characteristics of
solitons. The $B=1$ quantities as well as $B=2$ quantities for the torus 
calculated previously, are shown for comparison. \\}
\vspace{2mm}

As can be seen from {\bf Tables 1 and 2}, there are two ``islands" of 
stability for
baryon numbers considered here: at $B=4$, which is not unexpected, and for
$B=7$, and this is something new and unexpected. So far, this
property seems to be specific to the Skyrme model. The difference between
$\Theta_T$ and $\Theta_{T,3}$ is maximal for the toroidal $B=2$ configuration
and decreases with increasing $B$. It vanishes for $B=3$ and $7$ precisely.
The accuracy of calculations decreases with increasing baryon number. It is 
difficult to estimate it for such quantities as $\omega_{F,B}$ and $c_{F,B}$
where it depends also on the particular method of calculation - the rigid
oscillator model in our case.

The behaviour of static properties of multiskyrmions and flavour 
excitation frequencies shown in { \bf Table 1} is similar to that obtained in
\cite{22} for toroidal configurations with $B=2,3,4$.
We note that the flavour inertia $\Theta_{F,B}$
increases with $B$ almost proportionally to $B$. 
\section{Flavour excitation frequencies and $\sim 1/N_c$ zero mode corrections}
To quantize the solitons in their $SU(3)$ configuration space,  in the
spirit of the bound state approach to the description of strangeness
proposed in \cite{13}-\cite{14} and used in \cite{15}-\cite{17},
we consider the collective coordinate motion of the 
meson fields incorporated into the matrix $U$:
$$ U(r,t) = R(t) U_0(O(t)\vec{r}) R^{\dagger}(t), \qquad
 R(t) = A(t) S(t), \eqno (6) $$
where $U_0$ is the $SU(2)$ soliton embedded into $SU(3)$ in the usual way (into
the upper left hand  corner), $A(t) \in SU(2)$ describes $SU(2)$ rotations and 
$S(t) \in SU(3)$ describes 
rotations in the ``strange", ``charm" or ``bottom" directions 
and $O(t)$ describes rigid rotations in real space.
For definiteness we consider the extension of the $(u,d)$ $SU(2)$
Skyrme model in the $(u,d,s)$ direction, when $D$ is the field of $K$-mesons 
but it is clear that quite similar extensions can also be made
in the directions of charm or bottom. So

$$ S(t) = exp (i  {\cal D} (t)),  \qquad
 {\cal D} (t) = \sum_{a=4,...7} D_a(t) \lambda_a, \eqno (7) $$
where $\lambda_a$ are Gell-Mann matrices of the $(u,d,s)$, $(u,d,c)$ or $(u,d,b)$
$SU(3)$ groups. The $(u,d,c)$ and $(u,d,b)$ $SU(3)$ 
groups are quite analogous to
the $(u,d,s)$ one. For the $(u,d,c)$ group a simple redefiniton of 
hypercharge should be made. For the $(u,d,s)$ group,
 $D_4=(K^++ K^-)/\sqrt{2}$, $D_5=i(K^+-K^-)/\sqrt{2}$, etc.
For the $(u,d,c)$ group $D_4=(D^0 + \bar{D}^0)/\sqrt{2}$, etc.

The angular velocities of the isospin rotations $\vec{\omega}$ are defined
in the standard way \cite{9}:
$ A^{\dagger} \dot{A} =-i \vec{\omega} \vec{\tau}/2. $
We shall not consider here the usual space rotations in detail because the
corresponding momenta of inertia for $BS$ are much greater than the
isospin
momenta of inertia, and for the lowest possible values of angular momentum
$J$, the 
corresponding quantum correction is either exactly 
zero (for even $B$), or small, see also formulas $(17)$ and $(21)$ below.

The field $D$ is small in magnitude. In fact, it is, at least,
 of order $1/\sqrt{N_c}$, where $N_c$ is the number of colours in $QCD$,
see Eq. $(14)$.
Therefore, the expansion of the matrix $S$ in $D$ can be made safely. 

The mass term of the Lagrangian $(1)$ can be calculated exactly, without
expansion in the powers of the field $D$, because the matrix $S$ is given by
 $ S=1-i{\cal D} \; sind/ d
-{\cal D}^2 (1-cosd)/ d^2 $ with $d^2= Tr {\cal D}^2 $.
We find that
$$\Delta{\cal L}_M=-\frac{F_D^2m_D^2-F_\pi^2m_\pi^2}{4} (1-c_f)s_d^2 \eqno (8)$$
The expansion of this term can be done easily up to any order in $d$.
The comparison of this expression with $\Delta L_M$, within the collective
coordinate approach of the quantization of $SU(2)$ solitons in the $SU(3)$
configuration space \cite{10}-\cite{12}, allows us to establish
 the relation $sin^2 d =sin^2 \nu$,
where $\nu$ is the angle of the $\lambda_4$ rotation, or the rotation into
the ``strange" (``charm", ``bottom" ) direction.
 
After some calculations we find that the Lagrangian  
of the model, to the lowest order in the field $D$, can be written as
$$ L=-M_{cl,B}+4\Theta_{F,B} \dot{D}^{\dagger}\dot{D}-\biggl[\Gamma_B \biggl(
\frac{F_D^2}{F_{\pi}^2} m_D^2-m_{\pi}^2 \biggr)+ \tilde{\Gamma}_B(F_D^2-F_\pi^2)
\biggr] D^{\dagger}D -
 i{N_cB \over 2}(D^{\dagger}\dot{D}-\dot{D}^{\dagger}D). \eqno(9)$$
Here and below $D$ is the doublet $K^+, \, K^0$ ($D^0, \, 
D^-$, or $B^+,\,B^0$): $d^2=Tr {\cal D}^2=2D^\dagger D$.
We have kept the standard notation for the moment of inertia of the
rotation into the ``flavour" direction $\Theta_F$ for $\Theta_c, \,
\Theta_b$ or $\Theta_s$ \cite{10}-\cite{12}; different notations are used in 
\cite{15,16} (the index $c$ denotes the 
charm quantum number, except in $N_c$). The contribution proportional to
$\tilde{\Gamma}_B$ is suppressed in comparison with the term $\sim \Gamma$
by a small factor $\sim (F_D^2-F_\pi^2)/m_D^2$, and is more 
important for strangeness.

The term proportional to $N_cB$ in $(9)$ arises from the Wess-Zumino term
in the action and is responsible for the difference of
the excitation energies of strangeness and antistrangeness 
(flavour and antiflavour in the general case) \cite{13}-\cite{16}.

Following the canonical quantization procedure the Hamiltonian of the 
system, including the terms of the order 
of $N_c^0$, takes the form \cite{15,16}:
$$H_B=M_{cl,B} + {1 \over 4\Theta_{F,B}} \Pi^{\dagger}\Pi +
\biggl[\Gamma_B 
\bar{m}^2_D+\tilde{\Gamma}_B(F_D^2-F_\pi^2)+\frac{N_c^2B^2}{16\Theta_{F,B}} 
\biggr] D^{\dagger}D +i {N_cB \over 8\Theta_{F,B}}
(D^{\dagger} \Pi- \Pi^{\dagger} D), \eqno (10) $$
where $\bar{m}_D^2 = (F_D^2/F_{\pi}^2) m_D^2-m_\pi^2$.
The momentum $\Pi$ is canonically conjugate to variable $D$ (see Eq.$(18)$
below).
Eq. $(10)$ describes an oscillator-type motion of the field $D$ 
in the background formed 
by the $(u,d)$ $SU(2)$ soliton. After the diagonalization, which can be done
explicitly following \cite{15,16}, the normal-ordered Hamiltonian can be 
written as
$$H_B= M_{cl,B} + \omega_{F,B} a^{\dagger} a + \bar{\omega}_{F,B} b^{\dagger} b
 + O(1/N_c) \eqno (11) $$
with $a^\dagger$, $b^\dagger$ being the operators of creation of strangeness
(i.e., antikaons) and antistrangeness
(flavour and antiflavour) quantum number, $\omega_{F,B}$ and 
$\bar{\omega}_{F,B}$ being the 
frequencies of flavour (antiflavour) excitations. $D$ and $\Pi$ are connected
with $a$ and $b$ in the following way \cite{15,16}:
$$ D^i= \frac{1}{\sqrt{N_cB\mu_{F,B}}}(b^i+a^{\dagger i}), \qquad
\Pi^i = \frac{\sqrt{N_cB\mu_{F,B}}}{2i}(b^i - a^{\dagger i}) \eqno (12) $$
with
$$ \mu_{F,B} =[ 1 + 16 (\bar{m}_D^2 \Gamma_B+(F_D^2-F_\pi^2)\tilde{\Gamma}_B)
 \Theta_{F,B}/ (N_cB)^2 ]^{1/2}. \eqno (13) $$
For the lowest states the values of $D$ are small:
$$D \sim\bigl[16\Gamma_B\Theta_{F,B}\bar{m}_D^2+N_c^2B^2 \bigr]^{-1/4},\eqno (14) $$
and increase, with increasing flavour number $|F|$, like $(2|F|+1)^{1/2}$.
As  was noted in \cite{16}, deviations of the field $D$ from the vacuum 
decrease with increasing mass $m_D$, as well as with increasing number of 
colours $N_c$, and the method works for any $m_D$ 
(and also for charm and bottom quantum numbers).

The excitation frequencies $\omega$ and $\bar{\omega}$ are:
$$ \omega_{F,B} = \frac{N_cB}{8\Theta_{F,B}} ( \mu_{F,B} -1 ), \qquad
 \bar{\omega}_{F,B} = \frac{N_cB}{8\Theta_{F,B}} ( \mu_{F,B} +1 ) .\eqno (15)$$
As  was observed in \cite{17}, the difference 
$\bar{\omega}_{F,B}-\omega_{F,B}=N_cB/(4\Theta_{F,B})$ coincides, to the 
leading order in $N_c$, with the expression obtained in the collective coordinate
 approach \cite{18,19}. At large $m_D$ the  $\mu_{F,B} \simeq 4\bar{m}_D
 (\Gamma_B\Theta_{F,B})^{1/2}/(N_cB)$ 
and for the difference $\omega_{F,1}-\omega_{F,B}$ we obtain $(N_c=3)$:
$$\omega_{F,1}-\omega_{F,B} \simeq \frac{\bar{m}}{2} \biggl[\biggl
(\frac{\Gamma_1}
{\Theta_{F,1}}\biggr)^{1/2}-\biggl(\frac{\Gamma_B}{\Theta_{F,B}}
\biggr)^{1/2}\biggr] +{3 \over 8}\biggl( {B \over \Theta_{F,B}} -
{1 \over \Theta_{F,1}} \biggr). \eqno(16)$$
Obviously, at large $m_D$, the first term in $(16)$ dominates and is 
positive if $\Gamma_1/
\Theta_{F,1} \geq \Gamma_B/ \Theta_{F,B}$. This is confirmed by looking at 
{\bf Table 1}. Note also that the bracket in the first term in $(16)$ does
not depend on the parameters of the model if the background $SU(2)$
soliton is calculated in the chirally symmetrical limit: both $\Gamma$ and
$\Theta$
scale like $ \sim 1/(F_\pi e^3)$. In a realistic case when the physical pion
mass is included in $(3)$ there is some weak dependence on the parameters
of the model.

The $FSB$ in the flavour decay constants, i.e. the fact that $F_K/F_\pi 
\simeq 1.22$ and $F_D/F_\pi=1.7 \pm 0.2$, should be taken into account.
 In the Skyrme model this fact leads to the increase of the flavour excitation 
frequencies which changes the spectra of flavoured $(c,\,b)$ baryons and
puts them in a better agreement 
with the data \cite{20,21}. It also leads 
 to some changes of the total binding energies of $BS$ 
\cite{17}. This is partly due to the large contribution of the Skyrme term
to the flavour moment of inertia $\Theta_F$. Note, that in \cite{16} the 
$FSB$ in strangeness decay constant was not taken into account, and this led to
underestimation of the strangeness excitation energies. Heavy 
flavours $(c,b)$ have not been considered in these papers.

The terms of the order of $N_c^{-1}$ in the Hamiltonian, which  depend
 on the angular velocities of rotations in the isospin and the usual space
 and which 
describe the zero-mode contributions, are not crucial but important
for the numerical estimates of the spectra of baryonic systems.
To calculate them one should first obtain the Lagrangian of $BS$ including
all the terms upto $O(1/N_c)$. The Lagrangian can be written in
a compact form as:
$$L \simeq-M_{cl}+4\Theta_{F,B} \bigl[\dot{D}^{\dagger}\dot{D} \bigl(1-
{d^2 \over 3} \bigr) - {2 \over 3}(D^\dagger \dot{D} \dot{D}^\dagger D-
(D^\dagger\dot{D})^2-(\dot{D}^\dagger D)^2\bigr)\bigr] +
2\Theta_{F,B}(\vec{\omega}\vec{\beta}) +{\Theta_{T,B} \over2}
(\vec{\omega}-\vec{\beta})^2- $$
$$ -[\Gamma_B\tilde{m}_D^2+(F_D^2-F_\pi^2)
\tilde{\Gamma}_B] D^\dagger D \bigl(1 - {d^2 \over 3}\bigr) +i {N_cB \over 3}
\bigl(1- {d^2 \over 3}\bigr)(\dot{D}^\dagger D- D^\dagger \dot{D}) - {N_cB
\over 2} \vec{\omega} D^\dagger \vec{\tau} D,
 \eqno(17a) $$
where  $d^2=2D^\dagger D$ and
$$ \vec{\beta}=-i(\dot{D}^\dagger \vec{\tau} D - D^\dagger 
\vec{\tau} \dot{D}).$$ 
For the axially symmetrical configurations, like the $B=2$ torus, the term
$ \Theta_{T,B}(\omega_3-\beta_3)^2/2$ in $(17a)$ should be substituted by
$$ \delta L= \frac{\Theta_{3,B}}{2}(\omega_3- n \Omega_3 -\beta_3)^2 +
\frac{\Theta_{J,B}}{2}(\Omega_1^2+\Omega_2^2), \eqno (17b) $$
where 
$\Omega_i$ are the components of the angular velocities of rotation in
the usual space, $\dot{O}_{in}O_{kn}=\epsilon_{ikl}\Omega_l$.
Taking into account the terms $\sim 1/N_c$ we find that the canonical 
variable $\Pi$ conjugate to $D$  is
$$ \Pi =\frac{\partial L}{\partial\dot{D}^{\dagger}}=
4\Theta_{F,B} \bigl[\dot{D}\bigl(1-{d^2 \over 3} \bigr)-
{2 \over 3} D^{\dagger}\dot{D} \, D+{4 \over 3}\dot{D}^{\dagger}D \, D\bigr]
+i(\Theta_{T,B}-2\Theta_{F,B})\vec{\omega}\vec{\tau} D-
i\Theta_{T,B}\vec{\beta}\vec{\tau}D +i {N_cB \over 2}\bigl(1- 
{d^2 \over 3} \bigr) \, D . \eqno (18)$$

>From $(17a)$ the body-fixed isospin operator is:
$$ \vec{I}\sp{bf}=\partial L / \partial
\vec{\omega}=\Theta_{T,B}\vec{\omega}+(2\Theta_{F,B}-\Theta_{T,B})
\vec{\beta} - {N_cB \over 2} D^\dagger \vec{\tau} D.   \eqno (19) $$
Using the identities :
$$ -i \vec{\beta} \vec{\tau}D=2 D^\dagger D\dot{D}-(\dot{D}^\dagger
D+D^\dagger \dot{D}) D \eqno (20 a) $$
and
$$ \vec{\beta}^2 = 4 D^\dagger D \dot{D}^\dagger\dot{D}-(\dot{D}^\dagger D
+D^\dagger \dot{D})^2 \eqno (20b) $$
we find  that
the $\sim 1/N_c$ zero mode quantum corrections to the energies of skyrmions
can be estimated \cite{15,16} as:
$$\Delta E_{1/N_c} = {1 \over 2\Theta_{T,B}}\bigl[c_{F,B} T_r(T_r+1)+
(1-c_{F,B})I(I+1) + (\bar{c}_{F,B}-c_{F,B})I_F(I_F+1) \bigr], \eqno(21a) $$
where $I=I^{bf}$ is the value of the isospin of the baryon or $BS$, which
can be written also as:
$$ \vec{I}^{bf}= \Theta_T \vec{\omega}+\bigl(1 
-\frac{\Theta_T}{2\Theta_F}\bigr) \vec{I}_F -\frac{N_cB \Theta_T}{4\Theta_F}
D^{\dagger}\vec{\tau}D \eqno (22) $$
with the operator $\hat{\vec{I}}_F = (b^{\dagger}\vec{\tau}b - 
a^T\vec{\tau}a^{\dagger T})/2$.

$T_r$ is the quantity analogous to the
``right" isospin $T_r$, in the collective coordinate  approach \cite{10,18},
and $\vec{T_r}=\vec{I}^{bf}-\vec{I_F}$. The hyperfine structure constants
          $c_{F,B}$ and $\bar{c}_{F,B}$ are defined by relations:
$$ 1-  c_{F,B}=\frac{\Theta_{T,B}}{2\Theta_{F,B}\mu_{F,B}}(\mu_{F,B}-1), 
\qquad
1 - \bar{c}_{F,B}=\frac{\Theta_{T,B}}{\Theta_{F,B}(\mu_{F,B})^2}(\mu_{F,B}-1).
 \eqno(23)$$ 
          To take into account the usual space rotations the $J$-dependent
          terms should be added to $(21a)$. For the
          axially-symmetrical configurations, like the $B=2$ torus,
     they  are equal to \cite{18,16}:
$$ \Delta E_{1/N_c} = \biggl(\frac{1}{2n^2\Theta_{3,B}} - 
\frac{1}{2n^2\Theta_{T,B}} - \frac{1}{2\Theta_{J,B}}\biggr) (J_3^{bf})^2 
+ \frac{J(J+1)}{2\Theta_{J,B}}, \eqno (21b) $$ with 
$\Theta_{J,B}$ being the moment of inertia corresponding to the usual space
rotations - orbital moment of inertia, which is known to increase with increasing $B$-number almost proportionally to $B^2$ \cite{17,23}.
For such configurations the body-fixed $3$-$d$ component of the angular 
          momentum
$J^{bf}_3$ and the nonstrange part of the $3$-$d$ component of the isospin
(also body-fixed) are connected by the relation $J^{bf}_3 = - n T^{bf}_{r,3}$
( see, e.g. \cite{18,16} and references therein).
Realistic cases of multiskyrmions are intermediate between the case of 
incoherence of usual space and isospace rotations and the complete coherence, 
as in $(21b)$ for the rotation relative to the axis of axial symmetry.
However, the $J$-dependent terms of the type $(21b)$ are cancelled 
mostly in the differences of energies of states which belong
to the same $SU(3)$ multiplet, {\it i.e.} when they have the same values of
$J,\, (p,q) $ and $T_{r,3}$.
          
In the case of antiflavour excitations we obtain the same formulas, with 
the substitution $\omega \to \bar{\omega}$ and $ \mu \to -\mu $
in $(23)$. For example,
$$ \bar{c}_{\bar{F},B}=1 + \frac{\Theta_{T,B}}{\Theta_{F,B}\mu^2_{F,B}}
(\mu_{F,B}+1). \eqno (24) $$

The excitation energies for antiflavours are close to $\sim 0.59 Gev$ for 
antistrangeness, $\sim 1.75 \, Gev$ for anticharm and to $\sim 4.95 \, Gev$
for antibottom. However, these numbers should be considered as lower 
bounds only since to calculate them we have used a simplified version of the bound state soliton model.
\section{Estimates of the spectra of multibaryons with strangeness, charm or
bottom}
   In the bound state soliton model, and in its rigid oscillator version as
well, the states predicted do not correspond originally to
the definite $SU(3)$ or $SU(4)$ representations. How this can be remedied
 was shown in \cite{16}; see also Eq. $(26)-(29)$ below. The quantization
condition $(p+2q)/3=B$ \cite{10}, for arbitrary $N_c$,  changes to 
$(p+2q)=N_cB+3n_{q\bar{q}}$, where $n_{q\bar{q}}$ is the number of
additional valent quark-antiquark pairs present in the quantized 
states \cite{18}. 
For example, the state with $B=1$, $|F|=1$, $I=0$ and
$n_{q\bar{q}}=0$ should belong to the octet of $(u,d,s)$, or $(u,d,c)$,
 $SU(3)$ group, if $N_c=3$; see also \cite{16}. The state with $B=2$,
$|F|=2$ and $I=0$ should belong to the $27$-plet of the 
corresponding group, etc.
The states having antiflavour quantum number, i.e. positive strangeness or 
bottom quantum number or negative charm should have the number of additional
quark-antiquark pairs $n_{q\bar{q}} \geq |F|$ \cite{18}.
If $\Theta_F \rightarrow \infty$, Eqs. $(21)$ go
 over into the expression obtained within the
collective coordinate  approach \cite{10,17}. In a realistic case, with 
$\Theta_T/\Theta_F^{(0)} \sim 2 \,-\, 2.7$, the structure of $(21)$ is more 
complicated.

First we consider quantized states of $BS$ which belong to the lowest
possible $SU(3)$ irreps $(p,q)$ for each value of the baryon number, 
$p+2q=3B$: $p=0, \; q=3B/2$ for
even $B$, and $p=1, \; q=(3B-1)/2$ for odd $B$. For $B=3,\, 5$ and $7$ they 
are $\bar{35}, \, \bar{80}$ and $\bar{143}$-plets,
for $B=2, \, 4, \, 6$ and $8$ - $\bar{10}$, $\bar{28}$,  $\bar{55}$ and
$\bar{91}$-plets.
Since we are interested in the lowest energy states, we discuss here the
baryonic systems with the lowest allowed angular momentum, {\it ie} $J=0$,
 for $B=2$,
$4, \; 6$ and $8$. For odd $B$ the quantization of $BS$ 
encounters some difficulties
(see  \cite{23}),
but the correction to the energy of quantized states due to the nonzero angular
momentum is small and decreases with increasing $B$ since the corresponding
moment of inertia increases proportionally to $\sim B^2$ \cite{22,23}. 
Moreover, the $J$-dependent correction to the energy cancels in the 
differences of energies of flavoured and flavourless states.

For the energy difference between the state with flavour $F$ belonging to the
$(p,q)$ irrep, and the ground state with $F=0$ and the same angular momentum
and $(p,q)$ we obtain:
$$ \Delta E_{B,F} = |F| \omega_{F,B} +
\frac{\mu_{F,B}-1}{4\mu_{F,B}\Theta_{F,B}} [I(I+1)-T_r(T_r+1)]
 + \frac{(\mu_{F,B}-1)(\mu_{F,B}-2)}{4\mu_{F,B}^2 \Theta_{F,B}}
I_F(I_F+1), \eqno (25) $$
where $T_r=p/2$ and usually $I_F=I-T_r$. Note that the moment of
inertia $\Theta_T$ does not enter the difference of energies $(25)$.
Obviously, for ``minimal" $BS$, i.e. those which do not contain additional
quark-antiquark pairs, 
$$ T_r \leq 3B/2. \eqno (26) $$
The maximal isospin carried by $|F|$ flavoured mesons  bound by $(u,d)$
solitons satisfies another obvious relation:
$$I_F = |F|/2. \eqno (27) $$
Simple arguments allow us also to get the following restrictions on
 the total isospin of $BS$:
$$|T_r-|F|/2| \leq I \leq T_r+|F|/2 \eqno (28) $$
and
$$ I \leq (3B-|F|)/2. \eqno (29) $$
The lowest of the two upper bounds should be taken
as the final upper bound.
It is easy to check that our bounds correspond to the known $SU(3)$
multiplets for each value of $T_r$.

For the $B=1$ case, the difference of masses within the octet of baryons, 
$\Lambda_F$ and nucleon, $\Sigma_F$ and $\Lambda_F$, is  
$$ \Delta M_{\Lambda_FN} = \omega_{F,1}- \frac{3(1-\bar{c}_{F,1})}
{8\Theta_{T,1}}=\omega_{F,1}-\frac{3(\mu_{F,1}-1)}{8\mu_{F,1}^2\Theta_{F,1}},
\qquad \Delta M_{\Sigma_F \Lambda_F}= \frac{(1-c_{F,1})}
{\Theta_{T,1}}=\frac{\mu_{F,1}-1}{2\mu_{F,1}\Theta_{F,1}}. \eqno (30) $$

Clearly, there are cancellations in Eq. $(25)$ - the binding energy
differencies of multiskyrmions.
For states with maximal isospin $I=T_r+|F|/2$ 
the energy difference can be simplified to:
$$\Delta E_{B,F}=|F|\biggl[\omega_{F,B}+ T_r
\frac{\mu_{F,B}-1}{4\mu_{F,B}\Theta_{F,B}}
+\frac{(|F|+2)}{8\Theta_{F,B}} 
\frac{(\mu_{F,B}-1)^2}{\mu_{F,B}^2} \biggr]. \eqno (31) $$
For even $B$ $T_r=0$, 
for odd $B$ we should take $T_r=1/2$ for the lowest $SU(3)$ irreps.

\vspace{2mm}
\begin{center}
\begin{tabular}{|l|l|l|l||l|l|l||l|l|l|}
\hline
 $B$ & $\Delta \epsilon_{s=-1}$&
$\Delta\epsilon_{c=1}$ &$\Delta \epsilon_{b=-1}$& 
$\Delta \epsilon_{s=-2}$&
$\Delta\epsilon_{c=2}$ &$\Delta \epsilon_{b=-2}$&
$\Delta \epsilon_{s=-3}$&
$\Delta\epsilon_{c=3}$ &$\Delta \epsilon_{s=-4}$ \\
\hline
$2$&$-0.047$&$-0.027$&$0.02$&
$-0.115$&$-0.088$&$0.02 $&$-0.205$&$-0.183$&$-0.316$ \\
\hline
$3$&$-0.042$&$-0.010$&$0.04$&
$-0.098$&$-0.040$&$0.06 $&$-0.168$&$-0.064$&$-0.252$ \\
\hline
$4$&$-0.020 $&$0.019$&$0.06$&
$-0.051$&$ 0.022$&$0.10 $&$-0.092$&$0.013$&$-0.144$ \\
\hline
$5$&$-0.027 $&$0.006$&$0.05$&
$-0.063$&$0.001$&$0.08 $&$-0.108$&$0.019$&$-0.160$ \\
\hline
$6$&$-0.019 $&$0.016$&$0.05$&
$-0.045$&$0.023$&$0.10 $&$-0.078$&$0.028$&$-0.117$ \\
\hline
$7$&$-0.016 $&$0.021$&$0.06$&
$-0.041$&$0.033$&$0.11 $&$-0.070$&$0.037$&$-0.105$ \\
\hline
$8$&$-0.017 $&$0.014$&$0.02$&
$-0.040$&$0.021$&$0.03 $&$-0.068$&$0.020$&$-0.100$ \\
\hline
\end{tabular}
\end{center}
\vspace{2mm}
{\baselineskip=10pt 
\tenrm
{\bf Table 2.} 
The binding energy differences $\Delta \epsilon_{s,c,b}$ are the changes
of binding energies of lowest $BS$ with flavour $s,\,c$ or $b$ and isospin 
$I=T_r+|F|/2$ in comparison
with usual $u,d$ nuclei, for the flavour numbers $S=-1,\, -2, \, -3$ and $-4$,
$c=1,\, 2$ and $3$, $b=-1$ and $-2$ (see Eq. (32)).\\}
\vspace{2mm}
          
It follows from $(30)$ and $(31)$ that when some nucleons are replaced by
 flavoured hyperons in $BS$ the binding energy of the system  changes by
$$\Delta \epsilon_{B,F}=|F|\biggl[\omega_{F,1}-\omega_{F,B} - \frac{3(
\mu_{F,1}-1)}{8\mu_{F,1}^2\Theta_{F,1}}- T_r
\frac{\mu_{F,B}-1}{4\mu_{F,B}\Theta_{F,B}}
-\frac{(|F|+2)}{8\Theta_{F,B}} \frac{(\mu_{F,B}-1)^2}{\mu_{F,B}^2} \biggr].
\eqno (32) $$
For strangeness Eq. $(32)$ is negative indicating that stranglets should 
have binding energies smaller than those of nuclei, or can be unbound.
Since $\Theta_{F,B}$ and $\Theta_{T,B}$ increase with increasing $B$ and
$m_D$
this leads to the increase of binding with increasing $B$ and mass of the
flavoured state, in agreement with \cite{17}. For charm and bottom Eq.
$(32)$ is positive for $B \geq 3$ or $4$.
It follows from {\bf Table 2} that dibaryons with strangeness or charm quantum
number are probably unbound, but those with $b=-1$ or $b=-2$ could be bound.
The multibaryons with $B \geq 4$ and $S=-1$ can be bound, as well as 
multibaryons with $c=1,\, 2$ or $3$, or bottom $b=-1,\, -2$.

Had the momenta of inertia of $BS$ at small values of $B$ been
proportional
to the baryon number $B$, then the values of $\mu$, excitation frequencies
$\omega_F$ and coefficients $c$ would not have 
depended on $B$ at all. In this case the binding energy would have 
consisted only of its classical part and a contribution
from zero modes; the difference of $\omega$'s would have been absent in
this case.

The nuclear fragments with sufficiently large values of strangeness (or bottom)
may have been found in experiments as fragments with negative charge $Q$, 
according to the
well known relation, $Q=T_3+(B+S)/2$ (similarly for the bottom number).
Recently one event of a long lived nuclear fragment with mass about
$7.4 Gev$ was reported in \cite{24}. Using the above formulas it is
not difficult to establish that this fragment may be the state with $B=-S=6$, or $B=7$
and strangeness $S=-4$, or $-3$, see also {\bf Table 3} below. Greater values of
strangeness are not excluded
since the method used here overestimates the flavour excitation energies,
especially for smaller baryon numbers and for the strangeness quantum number.

Another case of interest involves considering the $BS$ with isospin $I=0$.
In this case $I_F=T_r=|F|/2$ and so such states do not belong to the lowest 
possible $SU(3)$ multiplet for each value of $B$ (except for the case
$|F|=1$). For the energy difference between this 
state and a flavourless state belonging to the same $SU(3)$ irrep it is easy
to obtain:
$$\Delta E_{B,F}=|F|\biggl[\omega_{F,B}-
\frac{(|F|+2)}{8\Theta_{F,B}} \frac{(\mu_{F,B}-1)}{\mu_{F,B}^2} \biggr].
\eqno (33) $$
For the difference of
binding energies  of such a state and the ground $(u,d)$ state 
with lowest values $(p^{min},q^{min})$ we have the following estimate:
$$\Delta \epsilon_{B,F}=|F|\biggl[\omega_{F,1}-\omega_{F,B}-
\frac{3(\mu_{F,1}-1)}{8\Theta_{F,1}\mu_{F,1}^2}
+\frac{(|F|+2)}{8\Theta_{F,B}} \frac{(\mu_{F,B}-1)}{\mu_{F,B}^2} \biggr]
-{1 \over 2 \Theta_{T,B}} [|F|(|F|+2)/4- T_r^{min}(T_r^{min}+1)], \eqno (34)
$$
where $T_r^{min}=0$, or $1/2$.
Using this formula we find  the values given in 
{\bf Table 3.} For example, the $B=2$, $|F|=2$ state discussed previously 
in \cite{18} and later in \cite{16} belongs to the $27$-plet of 
the corresponding  $SU(3)$ group. In the case of strangeness it has appeared
already, probably, as a virtual level in the $\Lambda \Lambda$ system
\cite{24}.
\begin{center}
\begin{tabular}{|l|l|l|l||l|l|l||l|l|l|}
\hline
 $B$  &$\Delta \epsilon_{s=-1}$& $\Delta \epsilon_{c=1}$& $\Delta\epsilon_{b=-1}$
  &$\Delta \epsilon_{s=-2}$& $\Delta\epsilon_{c=2}$ & $\Delta \epsilon_{b=-2} $
  &$\Delta \epsilon_{s=-3}$& $\Delta \epsilon_{c=3}$ & $\Delta \epsilon_{b=-3} $\\
\hline
$2$& $-$&$-$  & $-$  & $-0.075$&$-0.029$&$0.02$&$-$ & $-$ & $-$\\
\hline
$3$&$0.000$&$0.034$&$0.07 $&$- $&$ -$  & $-$&$-0.083$&$0.002$&$0.09$ \\
\hline
$4$&$- $&$- $&$- $&$-0.047$&$0.030$  &$0.09$& $-$ & $-$ & $-$   \\
\hline
$5$&$-0.003 $&$0.032 $&$0.06 $&$- $&$-$  &$- $&$-0.060$&$0.035$&$0.12$ \\
\hline
$6$&$- $&$- $&$- $&$-0.044$&$ 0.025$  &$0.09$&$-$ & $-$ & $-$  \\
\hline
$7$&$ 0.000$&$0.040$&$0.07$&$- $&$- $  &$ -$&$-0.042$&$0.068$&$0.15$ \\
\hline
$8$&$- $&$- $&$- $&$-0.039$  &$ 0.023$ &$0.03$& $-$ & $-$ & $-$   \\ 
\hline
\end{tabular}
\end{center}
{\baselineskip=10pt 
\tenrm
{\bf Table 3.} The binding energies differences of lowest flavoured $BS$ with
isospin $I=0$ and the ground state with the same value of $B$ and
$I=0$ or $I=1/2$, see Eq. $(34)$. 
The first $3$ columns are for $|F|=1$, the next $3$ columns -
for $|F|=2$, and the last 3 - for $|F|=3$. The state with the value of
flavour $|F|$ belongs to the $SU(3)$ multiplet with $T_r= |F|/2$. \\}

We can see from this {\bf Table} that the $B=7, \; S=-3$ state has a binding
energy  smaller than the $(u,d)$ nucleus by $42 \, MeV$, i.e. it can
still be stable with respect to the strong decay, if we take into account 
the uncertainty of our estimates (recall that the nucleus $^7Li$ has the total
binding energy $ 39 \, Mev$.) The state with isospin equal to $2$ -
maximal value for $S=-3$ within the $(1,10)$ $SU(3)$ multiplet - has
somewhat greater binding energy,  see {\bf Table 2} .
The difference of energies of states with isospin $I=I^{max}=T_r+|F|/2$ and
$I=0$ and the same value of $F$ can be written as
$$E^{I^{max}}-E^{I=0}=
\Delta \epsilon_{B,F}^{I=0}-\Delta\epsilon_{B,F}^{I^{max}}
= \frac{|F|(|F|+2)}{8}
\biggl(\frac{\mu_{F,B}-1}{\Theta_{F,B}\,\mu_{F,B}} - {1 \over \Theta_{T,B}}
\biggr) -\frac{T_r(T_r+1)}{2\Theta_{T,B}}+T_r|F|\frac{\mu_{F,B}-1}{4
\Theta_{F,B}\,\mu_{F,B}} \biggr] \eqno (35a) $$
At large $|F|$ this is approximately given by
$$E^{I^{max}}-E^{I=0} \simeq
\frac{\vec{I}_F^2}{2\Theta_{T,B}} \bigl( 1\,-\, 2 c_{F,B} \bigr) \eqno (35b) $$
At large $B$ and $F$ the isoscalar states have smaller energy if $c_{F,B}
\leq 0.5$, see also {\bf Table 1}.

It is of interest to consider the case corresponding to the bulk
of ``flavoured matter", i.e. $p=q=B=|F|$. Such ``multilambda" states with
isospin equal to zero have the following value of $\Delta \epsilon$ for
$B \gg 1$:
$$\Delta \epsilon \simeq |F|\biggl[ \omega_{F,1}-\omega_{F,B}
+\frac{|F|+2}{8}
\biggl(\frac{\mu_{F,B}-1}{\Theta_{F,B}\mu_{F,B}^2} - {1 \over \Theta_{T,B}}
\biggr) -\frac{3(\mu_{F,1}-1)}{8\mu_{F,1}^2\Theta_{F,1}} \biggr]. \eqno
(36) $$
At large $|F|$ the sign of this expression depends on the sign of the
difference $(\mu_{F,B}-1)/(\Theta_{F,B}\mu_{F,B}^2) - 1/\Theta_{T,B}$.
To draw the final conclusion we need the knowledge of the behaviour of the ratio
$\Theta_{T,B}/\Theta_{F,B}$ at large $B$.
 For heavy flavours, $c$ and $b$, $\mu_{F,B} \gg 1$ and second term in
$(35)$ is negative, unless $\Theta_{T,B} \sim \mu_{F,B} \Theta_{F,B}$
which is not realistic (we have usually $\mu_s \sim 3$, $\mu_c \sim 15$
and $\mu_b \sim 73$). So, for heavy flavours it is not possible to obtain,
in this way,
the bulk of flavoured matter as quantized coherent multiskyrmions.
Other possibilities remain to be investigated, e.g. flavoured skyrmion 
crystals.

As in the $B=1$ case \cite{26}, the absolute values of masses of
multiskyrmions are controlled by the poorly
known loop corrections to the classic mass, or the Casimir energy. 
 And as has been done for the $B=2$ states, \cite{18}, 
 the renormalization procedure is necessary to obtain physically reasonable
values of these masses. 
 As the binding energy of the deuteron is $30 \; MeV$ instead of the measured 
value $2.2 \; MeV$ we see that $\sim 30 \; MeV$ characterises the uncertainty of our 
approach \cite{17,18}. But this uncertainty cancels in the differences of
binding energies calculated in {\bf Tables 2,3}.
\section{Conclusions}
Using rational map ansaetze as starting configurations we have calculated
the static characteristics of bound skyrmions with baryon numbers up to $8$.
The excitation frequencies for different flavours - strangeness, charm and
bottom - have been estimated using a rigid
oscillator version of the bound state approach of the chiral soliton
models. One notes that, in comparison with strangeness,
this approach works even better for $c$ and $b$ flavours \cite{20,21}.
Our previous conclusion that $BS$ with charm and bottom have more chance
to be bound by strong interactions than strange $BS$ \cite{17} is 
reinforced by the present investigation. Estimates of the binding energy 
differences of flavoured and flavourless states have some uncertainty,
about few tens of $Mev$, but the tendency for charm and bottom to be bound
stronger than strangeness is very clear.

A natural question now arises as to how these results depend on the choice 
of the parameters of the model. A set of parameters, which has been 
used extensively in the literature,
is, e.g. the set introduced in \cite{9} where the masses of the nucleon and
the $\Delta$ isobar have been fitted in the massless case and with physical
pion mass, correspondingly. In view of the large negative contribution of
the loop corrections, or the  Casimir energy,
we feel that this choice of parameters cannot be taken too seriously. 
But calculations
show that our results hold for this choice too. The energies of the
flavour excitation are somewhat smaller, however: for example, for strangeness
$\omega_s= 255 \, MeV$  for $B=1$ and $249 \, MeV$ for $B=4$, and we take
$F_K/F_\pi=1.22$  $(F_\pi=108\,Mev, \;e=4.84)$. Similar
changes take place for charm and bottom, and the conclusion that charmed or
bottomed $BS$ have good chances to be stable against strong interactions
remains valid \cite{17}.   

It should be kept in mind that corrections of the order of $1/N_c^2$ 
can lead to some change of our results. For example, the flavour moment of 
inertia changes to \cite{18}
$$ \Theta_F \rightarrow \Theta_F^{(0)} - D^\dagger D \frac{F_D^2-F_\pi^2}{8}
\int (1-c_f)(2-c_f) d^3r, \eqno (37) $$
Decrease of the moment of inertia could lead to some increase of the zero-mode
quantum corrections.

The apparent drawback of our approach is
that the motion of the system into the ``strange", ``charm" or ``bottom" 
directions is considered independently from other motions. Consideration of
the $BS$ with ``mixed" flavours is  possible in principle, but its 
treatment would be more involved (see, e.g. \cite{27} where the 
collective coordinate approach to the quantization of $SU(n)$ skyrmions has
been investigated).

Our results  agree qualitatively with the results of \cite{28}
where the strangeness excitation frequencies had been calculated within the
bound state approach. The difference is, however, in the behaviour of 
excitation frequencies: we have found that they decrease when the 
baryon number increases from $B=1$ to $4$,
thus increasing the binding energy of the corresponding $BS$.
This behaviour seems to be quite natural: there is an attraction between
$K,\,D$ or $B$ meson field and $B=1$ nucleon, and the attraction of a meson
by $2,\,3$ etc. nucleons is greater.
Similar results hold for ordinary nuclei: the binding energy of
a deuteron is $2.22\, Mev$ only, for $B=3$ it is about $8\,Mev$, for $B=4$ it
is already $28\,Mev$, and saturation takes place soon.

There is a further difference between the rigid oscillator variant of the
bound state model we have used here and the collective coordinate  
approach of soliton models studied previously \cite{10}-\cite{12}.
In the collective coordinate approach involving zero modes of solitons with 
a rigid or a soft rotator variant of the model, the 
masses of baryons are usually considerably greater than in the bound state 
approach when the Casimir energies are not taken into account
 \cite{26,29}. One of the sources of this difference is the presence
of a term of order $N_c/ \Theta_F$ in the zero-mode contribution to 
the rotation energy, which is absent in the bound state  
model. Recently it has been shown  by Walliser, for the $B=1$ sector within the
$SU(3)$ symmetrical $(m_K=m_\pi)$ variant of the Skyrme model \cite{29},
that this large contribution is cancelled almost completely by the 
kaonic 1-loop correction to the  
zero-point Casimir energy which is of the same order of magnitude, $N_c^0$ 
\cite{29}. This correction has also been recently calculated within the
bound state approach to the Skyrme model \cite{30}.

The charmed baryonic systems with $B=3, \, 4$ were considered in \cite{31}
within a potential approach. The $B=3$ systems were found to be
very near the threshold and the $B=4$ system was found to be stable 
with respect to the strong decay, with a binding energy of $\sim 10 \, MeV$.

Experimental searches for the baryonic systems with flavour different from
$u$ and $d$ could shed more light on the dynamics of heavy flavours in
few-baryon systems. The negative charge fragment seen in the $NA52$ CERN
experiment \cite{25} could be explained in our approach as a quantized $B=7$
skyrmion with strangeness $S=-3$ or $-4$. The other possibility is $B=6$
and $S=-6$ or $-7$. The value of strangeness can be greater since the rigid 
oscillator version of the model we consider here overestimates the strangeness 
excitation energies.

The threshold for the charm production on a free nucleon is about $12 GeV$,
and for the double charm it is  $\sim 25.2 \, GeV$. For bottom, the threshold on
a nucleon is  $\sim 70$ $GeV$. However, for nuclei as  targets the
thresholds are much lower due to the two-step processes with mesons in
intermediate states and due to the normal Fermi-motion of nucleons inside the 
target nucleus (see, e.g. \cite{32}). Therefore, the production of
baryons or baryonic systems with charm and bottom should be feasable in 
proton accelerators with energies of several tens of $GeV$, as well as in
heavy ions collisions.

Let us finish by adding that a shortened and less complete version of 
this paper is available \cite{33}. The results obtained recently in \cite{34}
within the detailed version of the bound state approach are in fair agreement 
with our, but the binding obtained in \cite{34} is smaller than
what we have found here. It 
should be noted that, in difference from \cite{34}, we have used the 
empirical values of flavour decay
constants, taken into account the $1/N_c$ zero modes contributions to the
energy of multibaryons and have considered only the difference of binding 
energies
of flavoured $BS$ and the ground states where many of uncertainties cancel out.

We are indebted to B.E.Stern for help in numerical calculations and to
J.Madsen for informing us about the result of NA52 experiment \cite{25}.
This work has been supported by the UK PPARC grant: PPA/V/S/1999/00004.
 \\
{\elevenbf\noindent References}
\vglue 0.1cm

\end{document}